\documentclass[12pt]{iopart}

\usepackage{graphicx}
\usepackage[usenames]{color}
\newcommand{\beq}{\begin{equation}}
\newcommand{\eeq}{\end{equation}}
\newcommand{\beqa}{\begin{eqnarray}}
\newcommand{\eeqa}{\end{eqnarray}}
\newcommand{\ba}{\begin{array}}
\newcommand{\ea}{\end{array}}

\begin{document}
\title{BCS-BEC crossover in a trapped Fermi super-fluid using a
density-functional equation}
\author{S. K. Adhikari\footnote{adhikari@ift.unesp.br;
URL: www.ift.unesp.br/users/adhikari} }
\address{Instituto de
F\'{\i}sica Te\'orica, UNESP - Universidade Estadual Paulista,
01.140-070 S\~ao Paulo, S\~ao Paulo, Brazil\\
}

\begin{abstract} We derive a generalized time-dependent 
Galilean-invariant density-functional (DF) equation appropriate to study 
the stationary and non-stationary properties of a trapped Fermi 
super-fluid in the Bardeen-Cooper-Schrieffer (BCS) to Bose-Einstein 
condensation (BCS) crossover. This equation is equivalent to a quantum 
hydrodynamical equation for a Fermi super-fluid. The bulk chemical 
potential of this equation  has the
proper 
(model-independent) dependence on the Fermi-Fermi scattering length in the 
BCS and BEC limits. We apply this DF equation to the study of stationary 
density profile and size of a cigar-shaped Fermi super-fluid of $^6$Li 
atoms and the results are in good agreement with the experiment of 
Bartenstein {\it et al.} in the BCS-BEC crossover. We also apply the DF 
equation to the study of axial and radial breathing oscillation and our 
results for these frequencies are in good agreement with experiments in 
the BCS-BEC crossover.

\end{abstract}

\pacs{71.10.Ay, 03.75.Ss,  67.85.Lm }

\maketitle

\section{Introduction}

\label{I}

The  Bardeen-Cooper-Schrieffer (BCS) equation valid in the 
weak-coupling limit (Fermi-Fermi scattering length $a\to -0$)
of a  Fermi super-fluid (SF) has been 
fundamental for the study of low-temperature superconductors \cite{BK}. 
Later, a 
microscopic field-theoretic formulation of the Fermi SF has been 
proposed by Bogoliubov and de Gennes \cite{BK}. 
Eagles \cite{cross}
suggested the possibility 
of extending the study of the Fermi SF to the domain of stronger 
coupling which initiated the study of the Fermi SF in the BCS to 
Bose-Einstein condensation (BEC) crossover \cite{cross2}. 
As the strength of the attractive interaction between two spin-up and 
-down fermions in a (Cooper) pair forming the Fermi SF is increased
($a<0$) 
the simple BCS SF turns into a complex
Cooper-pair-induced strongly interacting SF  at unitarity 
($a\to \pm \infty$) and beyond ($a>0$)
these two fermions 
form a diatomic 
molecule (dimer)
and undergo BEC. In this domain of interaction, the 
properties of the SF can largely be described by mean-field 
bosonic equations and in the so called BEC limit as 
 $a\to +0$, the system is ideally described by the 
mean-field Gross-Pitaevskii (GP) equation \cite{GP} for dimers. 
Although the studies of the Fermi SF in the weak-coupling BCS ($a\to 
-0$) and BEC ($a\to +0$) limits as well as at unitarity ($a\to \pm 
\infty$) are well under control, its study in the full BCS-BEC 
crossover, as $a$ varies from $-0$ to +0 through $\pm\infty$ remains to 
be a challenging and formidable task \cite{rmp2}.

After the
first observation of a strongly interacting degenerate Fermi gas \cite{ohara},
recently, several experimental groups have observed and studied a trapped 
Fermi SF in 
the the BCS-BEC 
crossover composed of {\color{Blue}$^{40}$K \cite{exK} and $^6$Li 
\cite{exK2,exLi}}
atoms 
 under controlled conditions. 
The crossover  in the Fermi SF was attained by 
varying an external background magnetic field near a Feshbach 
resonance \cite{fesh}
which allows an experimental manipulation of the 
 S-wave scattering length $a$. The study of the Fermi SF in the 
BCS-BEC crossover has gained new impetus \cite{rmp2,bulgac,castin,DF2,exp}
after these experiments and the 
necessity of a proper theoretical formulation to study the trapped Fermi 
SF in the crossover cannot be overemphasized. Recently, we formulated 
\cite{AS2,AS3}
a 
density-functional (DF) equation \cite{DFT}
which
produced results \cite{adh1,adh2}
for energy of a spherically-symmetric 
trapped 
Fermi SF in the BCS-unitarity crossover in close agreement with Monte
Carlo calculations \cite{MC}. 
Here we  extend  that DF equation  to the whole 
BCS-BEC crossover  and 
apply it to the study of stationary and non-stationary problems of a 
cigar-shaped \cite{gor} Fermi SF.  Usually, the use of any 
microscopic formulation of a Fermi SF  
(including the Monte Carlo simulations)
in a large system as encountered in laboratory is extremely complicated 
numerically and/or analytically and the advantages of a DF approach to Fermi 
systems have long been realized \cite{DFT}. 
The DF approach with a  small number of 
degrees 
of freedom 
is reasonably easily  formulated and applied to large Fermi systems \cite{DFT}.

Before we detail the DF formulation for Fermi SF, we list the desired aspects and 
properties of the Fermi SF which are included in the present 
formulation. (i) A Fermi SF has been long described by 
a Galilean-invariant 
hydrodynamical formulation \cite{hydro}. By showing complete equivalence of the 
present DF equation with a quantum 
 hydrodynamical equation, we preserve Galilean invariance of the present 
equation and include proper gradient (surface) correction \cite{AS2,AS3} 
in the Thomas-Fermi approximation (or the local-density approximation) 
\cite{rmp2}. (Such surface correction was first introduced by von 
Weizs\"acker \cite{von} in the study of properties of large nuclei.) 
(ii) There are model-independent results \cite{uni,lee,huang} for energy 
per atom of a uniform Fermi SF near the BCS limit in terms of the 
scattering length $a$, as well as of a SF of dimers in the weak-coupling 
BEC limit, and these will be included in our formulation.  (iii) A 
theoretical formulation of the the uniform Fermi SF at unitarity is 
particularly interesting as the only available scale of length at 
unitarity is $n^{-1/3}$, where $n$ is the density, and all energies 
should have the universal form \cite{the1,the1x} $\sim \hbar^2 
n^{2/3}/m$, where $m$ is the mass of a pair. This aspect of the Fermi SF 
at unitarity is included in the present DF equation.  
(An interesting account of the universal behavior of the properties 
of a Fermi gas at unitarity has recently appeared \cite{scienceuni}.) 
There are other DF-type 
equations for a Fermi SF valid for the crossover \cite{kz,Sala,adhi} and 
also at unitarity \cite{bulgac}, but they fail on one or more accounts 
above. They all fail to satisfy criterion (i) above. The studies 
of  \cite{kz,Sala} do not satisfy criterion (ii). It is not easy to 
realize if any of these properties are violated in the approximate 
solution of the Bogoliubov equation as reported in  \cite{perali}. 
(The accurate solution of the Bogoliubov equation should not violate any 
of these properties.)


In Sec. \ref{II} we present and discuss the DF equation 
for Fermi SF
in the BCS-BEC crossover. 
In Sec \ref{III} we present the numerical results of the 
present formulation. In Sec \ref{IIIa} we discuss the stationary   
properties, such as size and linear density of a cigar-shaped   
Fermi SF, and compare our results with experiment \cite{prl}.
In  Sec \ref{IIIb} we discuss the breathing oscillation 
of a cigar-shaped Fermi SF and compare the results for frequencies 
of radial and axial oscillation with experiment \cite{altme,kinast,barten}. 
Finally, in Sec. 
 \ref{IV} we present some concluding remarks.

\section{DF equation
for a Fermi SF in the BCS-BEC crossover}

\label{II}

We consider a
Fermi SF composed of an equal number of fully paired 
spin-up and  spin-down
fermions.
In the present study we start with 
the following DF equation \cite{adh1,adh2}
 which  produced results for energy of a Fermi SF 
in a spherically-symmetric harmonic trap
in close agreement
with  Monte-Carlo (MC) calculations \cite{MC} in the BCS-unitarity 
crossover \cite{cross2}
\begin{eqnarray} \label{eq1}
\left[-\frac{\hbar^2}{2m}\nabla^2+U({\bf r})+\mu(n,a)-i\frac{\partial}{\partial
t}\right]\Psi({\bf r},t)=0, \end{eqnarray}
where $U$ is the trapping
potential, $m$
is the mass of a pair (twice the atomic mass), $\mu(n,a)$ is the bulk 
chemical potential defined by \cite{adh1} 
\begin{eqnarray}\label{eq2}
\frac{\mu(n,a)}{2E_F}= 1+\frac{\chi_1 x-\chi_2 x^2}
{1-\beta_1x+\beta_2 x^2}, \quad x<0,
 \end{eqnarray}
where  $n \equiv |\Psi|^2$ is the density of pairs, 
the gas parameter 
$x\equiv n^{1/3}a$, $E_F\equiv  \hbar^2(6\pi^2 n)^{2/3}/m $ is the Fermi 
energy of the uniform gas. 
The normalization condition of the DF wave function $\Psi$ 
is
$\int|\Psi|^2 d^3r=N$, where $N$
is the number of pairs.
The Fermi-Fermi scattering length $a$ 
is negative corresponding to attraction: $-0>a>-\infty$.  
The constants of  (\ref{eq2})
have the following values \cite{adh1}:
$\chi_1=4\pi 2^{1/3}/(3\pi^2)^{2/3}, \chi_2=2^{2/3}300 , \beta_1=2^{1/3}40,$ 
and $\beta_2 = 
\chi_2/(1-\xi)$, where $\xi$ is an universal constant (called 
Bertsch parameter)
which determines 
the value of the bulk chemical potential at unitarity \cite{the1,the1x,the2}: 
\begin{equation}\label{uni}
\lim_{a\to \pm \infty}
\mu(n,a)=2E_F \xi .  
\end{equation}
At unitarity the only length scale in a uniform Fermi SF
is
$n^{-1/3}$, and from dimensional argument the chemical potential 
 of the trapped SF fermions (and also the energy)
 have the above universal
form \cite{uni}.
The bulk chemical potential $\mu(n,a)$ of (\ref{eq2}), 
has the following leading terms of dilute uniform gas in the BCS limit
($a\to -0$), as 
obtained from the model-independent 
expression for energy per particle derived 
by Lee and Yang \cite{uni,lee}
\begin{equation}\label{ly}
\frac{\mu(n,a)}{2E_F}=1 + \frac{4\pi 2^{1/3}}{(3\pi^2)
^{2/3}} (n^{1/3}a)+...\;. 
\end{equation}

The value of the Bertsch parameter  $\xi$ of  (\ref{eq2})
can be 
obtained from experiment on a trapped Fermi SF \cite{rmp2}
or from the MC calculation of a uniform Fermi SF \cite{the1,the1x}.  
There have been theoretical \cite{the1,the1x,the2} and experimental \cite{exp} 
investigations which extracted the value of the constant $\xi$ for a 
Fermi SF, and now there is a consensus \cite{rmp2} that its value should 
not depend on the specific Fermi atoms used. The experimental extraction 
of this constant from an analysis of trapped Fermi SF has 
yielded slightly different values ranging from 0.3 to 0.5 
under 
different trapping conditions \cite{rmp2,adhi}, 
whereas MC calculation in a uniform 
SF have yielded \cite{the1}
the value 0.44. {{A recent 
 average over experimental evaluations  of the Bertsch parameter   
is {\color{Blue}$\sim 0.38$ \cite{jltp} within about 2$\%$}.  The values of this 
parameter  are obtained from energy-entropy
measurements of
the ground state, sound speed, and cloud size.}}
(The BCS mean-field \cite{BCS}
calculation yielded 
slightly different result: $\xi= 0.59$.)
The spread of the 
experimental value of $\xi$ under different trapping conditions seems to 
be well outside the range of experimental uncertainty.  Hence a question 
naturally arises. Whether the value of $\xi$ for a trapped Fermi SF 
should depend on trapping conditions and number of atoms. We explore this possibility 
in the present investigation.

As the Fermi-Fermi attraction is increased beyond 
unitarity so that the 
scattering length $a$ turns positive,   the Fermi pairs, each composed  
of a spin-up and a spin-down
fermion,  
are transformed into bound molecules (dimers) which undergo BEC. 
In the weak-coupling  limit of weak interaction
(BEC limit) the Fermi SF is described by the GP equation 
for dimers. A proper description of the BEC of dimers should be made 
in terms of the dimer-dimer scattering length $a_d$. 
Using the zero-range approximation
Petrov 
{\it et al.} \cite{petrov}
calculated the dimer-dimer scattering length
$a_d$ in terms of the Fermi-Fermi scattering length $a$ and obtained 
\begin{equation}
a_d=0.6 a,
\end{equation}
which we shall use in the present study. 

In the BEC-unitarity crossover, 
as the dimer-dimer scattering length $a_d$ varies from $+0$ to $+\infty$, 
we shall adopt a recent beyond mean-field
extension of the GP equation \cite{AS1}
for 
fundamental bosons valid in the 
weak-coupling limit to  unitarity. 
For a  BEC of dimers, 
the energy and chemical potential have the same
universal form  (\ref{uni}). (For fundamental bosons considered in 
 \cite{AS1}, at unitarity, the constant $\xi$ had a value different 
from $\xi=0.44$, e.g. $\xi=0.73$ extracted from  (16) of  \cite{cow}.)  
We now generalize the bulk chemical potential $\mu(n,a)$
of  (\ref{eq2}) to the full BCS-BEC crossover. 
This can be achieved if we take $\mu(n,a)$
of  (\ref{eq1}) in the BEC-unitarity crossover ($+\infty> a,a_d>+0$) 
as
\begin{eqnarray}\label{eq3}
\frac{\mu(n,a)}{2E_F}=  \frac{2\pi}{(6\pi^2)^{2/3}}
\frac{y+(1+\nu)\alpha y^{5/2}}
{1+\nu \alpha y^{3/2}+(1+\nu)\gamma y^{5/2}}, \; \; y>0,
\end{eqnarray}
where $y=(a_dn^{1/3}), \alpha=32/(3\sqrt\pi)$, $\gamma=2\pi \alpha/[(6
\pi^2)^{2/3}\xi]$.
{\color{Blue}For any value of $\nu$ the bulk chemical potentials of  (\ref{eq2})
and (\ref{eq3}) are continuous at unitarity.} 
The only free parameter  $\nu$ in this expression should be adjusted to 
have good overall agreement to observables in the BEC side of crossover.
A set of equations, similar to ~(\ref{eq1}) and (\ref{eq3}), for
fundamental bosons (and not for composite dimers) produced results
for energy \cite{AS2} of a trapped condensate in agreement with
MC calculations \cite{MCBOSE}.

The form (\ref{eq3}) for the 
bulk chemical potential has several desired properties. 
For any $\nu$, this chemical potential
for the Fermi SF of dimers  in the BEC-unitarity crossover, by construction,
has 
the following
two leading terms  of a dilute uniform Bose gas
as obtained 
from the model-independent 
expression for energy per particle  derived 
by Lee, Huang, and Yang \cite{lee,huang}
\begin{eqnarray} \label{huang}
{\mu(n,a)}=({4\pi\hbar^2 a_d n}/{m})
\left[1+\alpha (n^{1/3}a_d)^{3/2}+...\right].\label{exp}
\end{eqnarray}
Higher-order terms of
expansion (\ref{exp}) has also been considered \cite{bra}; the
lowest order term was derived by Lenz \cite{lenz}.

Although (\ref{eq2}) and (\ref{eq3}) refer to Cooper-pair induced BCS SF 
and the dimer BEC SF, respectively, these bulk chemical potentials 
should be interpreted differently. For the BCS SF 
{\color{Blue}the bulk chemical potential}
 originates from the 
kinetic energy of Fermi atoms put in different quantum orbitals 
consistent with the Pauli principle discounted for by the negative 
attractive energy due to atomic interaction. For the dimer SF 
{\color{Blue}the bulk chemical potential}
originates solely from the repulsive interaction energy among dimers.

Considering only
the lowest-order term of  expansion (\ref{exp}) in  (\ref{eq1}), 
appropriate in the
BEC limit as $a_d\to 0^+$, the dimers obey the usual
GP equation \cite{GP} with the nonlinear bulk chemical 
potential $\mu(n,a)=4\pi \hbar^2 a_d n/m.$
Considering the first two terms   of  expansion (\ref{exp}), in the
BEC limit, one obtains a  modified GP equation for dimers
written by Fabrocini and Polls \cite{polls}.
But as $a_d$ (and  $a$)
increases and diverges at unitarity, the nonlinear bulk chemical potential 
should
saturate to the finite universal value  of 
(\ref{uni}) and should 
not diverge like the nonlinear terms of the GP
equation and of the Fabrocini-Polls equation given by  (\ref{huang}). 
The chemical potential and energy should not diverge at
unitarity, as the interaction potential remains finite in this limit,
although the scattering length $a$ diverges. In the weak-coupling GP
limit, the scattering length serves as a faithful measure of
interaction. But as the scattering length 
 increases, it ceases to be a measure of
interaction. Similarly, the bulk chemical potential of  (\ref{eq2}) should 
saturate, and not diverge, at unitarity.

It has been shown that the DF GP equation (\ref{eq1})  is 
Galilean-invariant and 
equivalent
to the quantum hydrodynamic equations for dimers \cite{AS2,adhi}
\begin{eqnarray}\label{h1}
\frac{\partial n}{\partial t}+\nabla \cdot (n {\bf v})=0 \,,\\
m \frac{\partial {\bf v}}{\partial
t}+\nabla\biggr[-\frac{\hbar^2}{2m} \frac{\nabla^2 \sqrt n}{\sqrt
n}+\frac{mv^2}{2}+U+ \mu(n,a)\biggr]=0\,, \label{h2}
\end{eqnarray}
if in  (\ref{eq1}) we take 
\begin{eqnarray}
\Psi({\bf r},t) &=& \sqrt{n({\bf r},t)}\, e^{i\theta({\bf r},t)} \,,
\\
{\bf v} &=& \hbar\nabla\theta/m \,,\label{pv}
\end{eqnarray}
where $\bf v$ is velocity of hydrodynamic flow of Fermi SF pairs or 
dimers of mass $m$
and $\theta$ is the phase of the wave function. The quantum hydrodynamical 
equations  (\ref{h1}) and  (\ref{h2})
are 
new and 
the usual 
classical 
hydrodynamical equations \cite{rmp2} 
for Fermi SF can be obtained by setting $\hbar=0$ in  (\ref{h2}).

In brief, the desired qualities of the present formulation are the 
following: (i) Galilean invariance and complete equivalence to 
hydrodynamical equations,  
(ii) correct BCS and BEC limits given by (\ref{ly}) and (\ref{huang}), 
respectively. There have been previous attempts \cite{kz,Sala,adhi,perali} 
to 
formulate dynamical equations for a trapped Fermi SF valid in the 
BCS-BEC crossover. The formulations of  \cite{kz,Sala,adhi}
had a different  
mass in the phase-velocity relation (\ref{pv}) and hence led to an 
inappropriate gradient term in  (\ref{eq1}) violating Galilean 
invariance and destroying the equivalence with the quantum 
hydrodynamical 
equations. The formulations of   \cite{kz,Sala} used mathematical 
fitting functions for the bulk chemical potential which did not satisfy the 
weak-coupling limits (\ref{ly}) and (\ref{huang}).

\section{Numerical results}

\label{III}

To see how DF equation (\ref{eq1}) with  (\ref{eq2}) and 
(\ref{eq3}) work in practice, we apply it for the numerical 
study of a Fermi SF in an axially symmetric trap. In the first part of 
this investigation we study the density profile and root mean square (rms)
 size
of the Fermi $^6$Li SF for the trap parameters 
 and compare our numerical results with experiment 
\cite{prl}. 
In the second part we calculate the frequencies of radial and axial 
oscillations  of a cigar-shaped \cite{gor}
Fermi SF and compare the results 
with experiments \cite{altme,kinast,barten}. 

We solve the partial differential equation (\ref{eq1}) in the BCS-BEC 
crossover by discretizing with the semi-implicit Crank-Nicolson 
algorithm and employing imaginary time propagation for the study of 
stationary ground states. For the study of dynamics we use real time 
propagation. To this end we use the FORTRAN programs provided in  
\cite{murgi}. For discretization we use the space step 0.04 in both 
radial and axial  directions and a time step of 0.0004
(both steps expressed in harmonic oscillator units in axial direction). 

\subsection{Stationary Fermi SF}

\label{IIIa}

The experiment of  \cite{prl} was performed with $^6$Li atoms in the 
BCS-BEC crossover 
using a Feshbach resonance at  an external magnetic field of 834 G. The results 
reported there are with a radial frequency of $\omega_\rho=2\pi \times 640$ Hz and
an axial frequency of $\omega _z=2\pi \times (600B/kG+32)^{1/2}$ Hz, where $B$ 
is the external magnetic field in kG. The results of the experiment 
are classified by the magnetic field $B$, which has been varied near 
the Feshbach resonance at 834.15 G   to control  the scattering length and 
achieve the BCS-BEC crossover.  The  parametrization of the external 
magnetic field $B$ (G)  in terms of the Fermi-Fermi scattering length $a$
\cite{para}
\begin{eqnarray}\label{para}
 a=-1405a_0\left[ 1+\frac{300}{B-834.15}\right]\left[1+
\frac{B-834.15}{2500}\right],
\end{eqnarray} 
with $a_0$ the Bohr radius,  
has been used to calculate the scattering length.
The results reported are with $4\times 10 ^5$ atoms or with $N=2\times 10^5$ 
molecules.

\begin{figure}
\includegraphics[width=\linewidth]{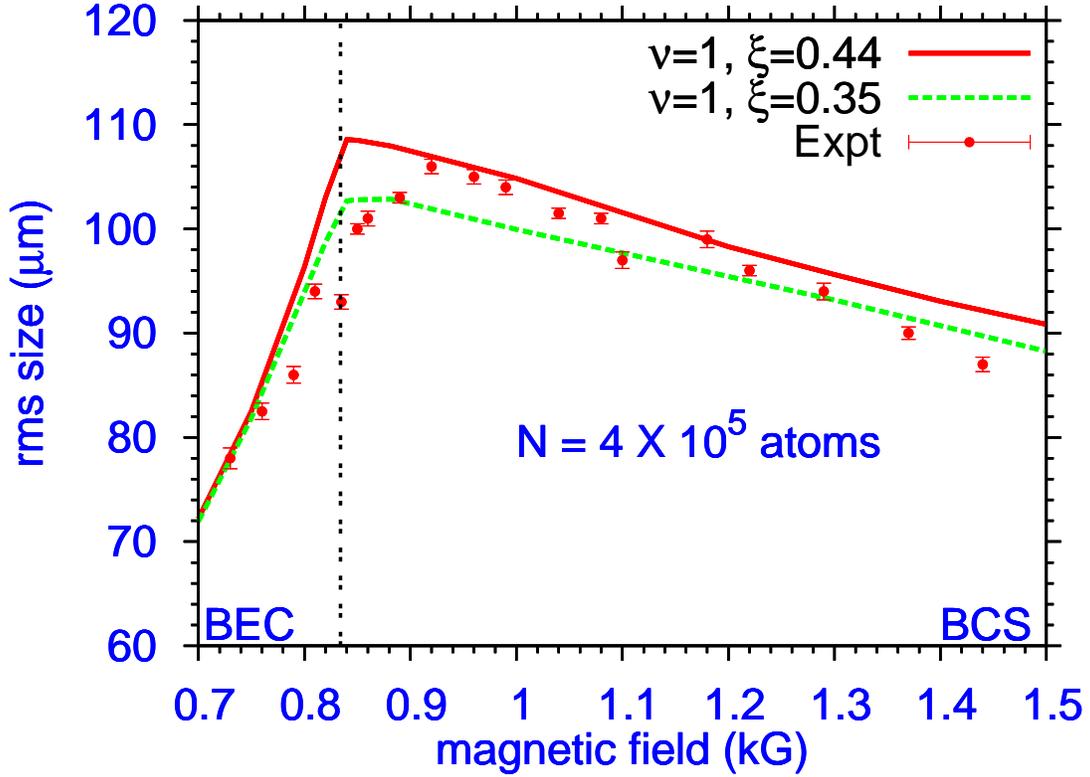}
\caption{(Color online) The rms size of the Fermi SF for different 
magnetic field  $B$  for $N=4\times 10^5$  $^6$Li atoms with the axially 
symmetric trap described in  \cite{prl}. Numerical  results for 
 $\xi=0.44, \nu=1$;  and 
$\xi=0.35, 
\nu=1$ are exhibited by lines and experimental results are shown with 
error bars. The vertical line at $B=0.834$ kG shows the position of the 
Feshbach resonance.
}
\label{fig2}
\end{figure}

The DF equations (\ref{eq1}) and (\ref{eq2}) valid in the BCS-unitarity 
crossover produced excellent results \cite{adh1,adh2}
for energy of a Fermi SF in a 
spherically symmetric trap, {\color{Blue}when compared to Monte Carlo calculations
\cite{MC},}  
at unitarity for 4 to 30 atoms and along the 
crossover for 4 and 8 atoms for $\xi=0.44$. In the BEC-unitarity 
crossover the relevant equations are (\ref{eq1}) and (\ref{eq3}) 
containing the free parameter $\nu$.

Now we study the rms size and density profile of the Fermi SF 
across the BCS-BEC crossover for different values of the magnetic field 
$B$, consequently for different values of scattering length. 
After some experimentation to fit the observed density profile and sizes 
\cite{prl} in the BEC side, we find that $\nu =1$ in  (\ref{eq3}) provides 
good overall fit and we use this value in all calculations here.  
At 
unitarity the Thomas-Fermi results for the rms size and density profile 
can be calculated analytically and the best fit to the experimental 
density profile is obtained for $\xi=0.27$ \cite{rmp2,prl}. 
This suggests the use of a 
smaller value of $\xi$ (than $\xi=0.44$) in the present analysis. Using 
the present DF equations we also find that 
$\xi \approx 
0.35$, 
rather than $\xi=0.44$, gives the best overall 
agreement with the 
experimental density profile for different $B$ (see Figs. \ref{fig3} below).
{\color{Blue}(The very small value for the Bertsch parameter 0.27 obtained in \cite{prl}
is probably due 
to systematic error in the experimental analysis using the inaccurate 
position of the Feshbach resonance at 850 G.  The later 
analysis  used the more precise  location of the Feshbach resonance 
at 834 G, giving a more reliable  value of the Bertsch parameter 
\cite{jltp}.)}  
In figure \ref{fig2} we plot the rms size vs. magnetic field $B$ for 
$\xi=0.44 $ and 0.35 together with the experimental results. 
From the 
general trend of the results exhibited in figure \ref{fig2}, it seems that 
$\xi=0.35$ gives good  agreement with experiment.

\begin{figure}
\begin{center}
\includegraphics[width=.49\linewidth]{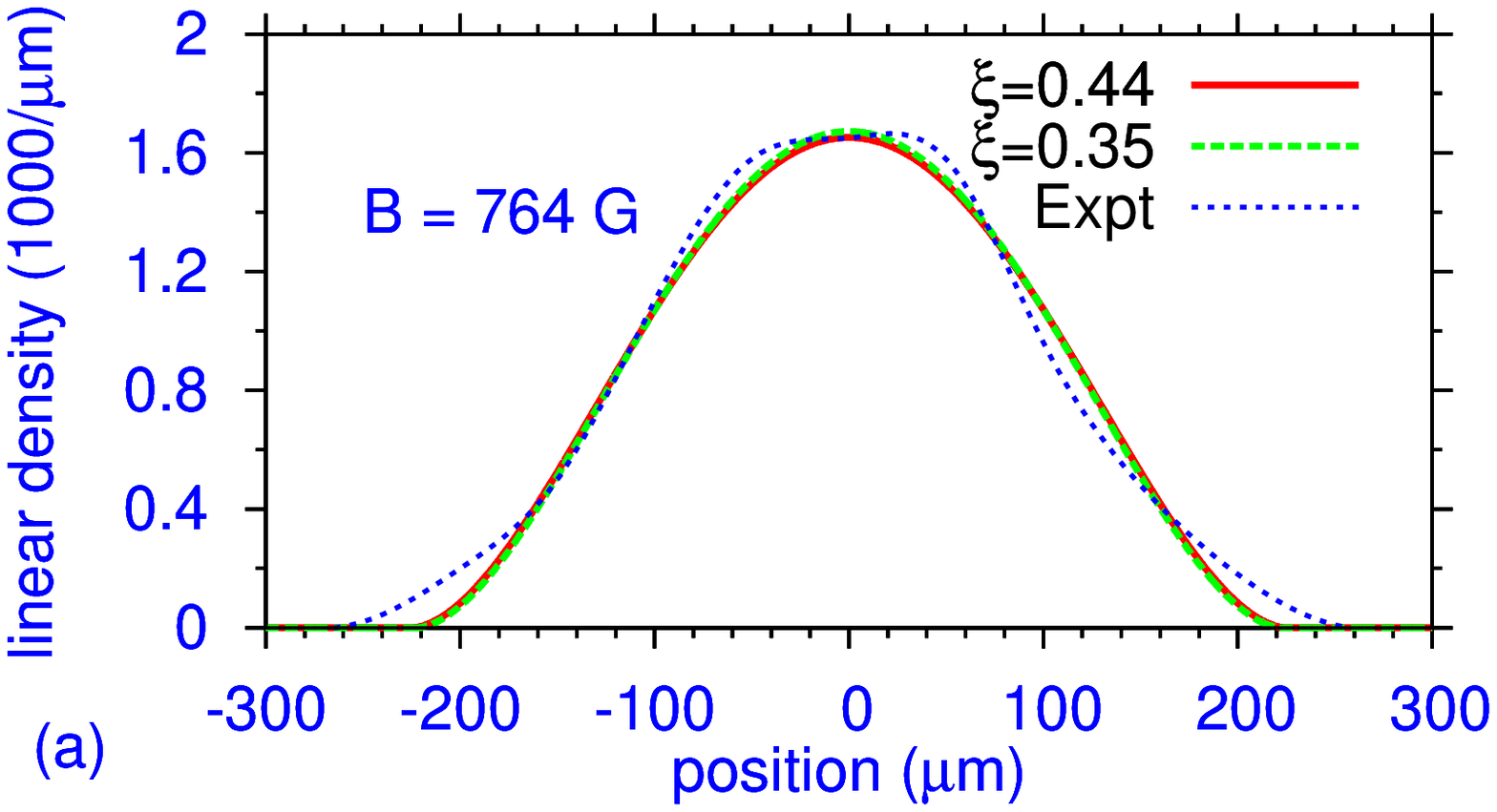}
\includegraphics[width=.49\linewidth]{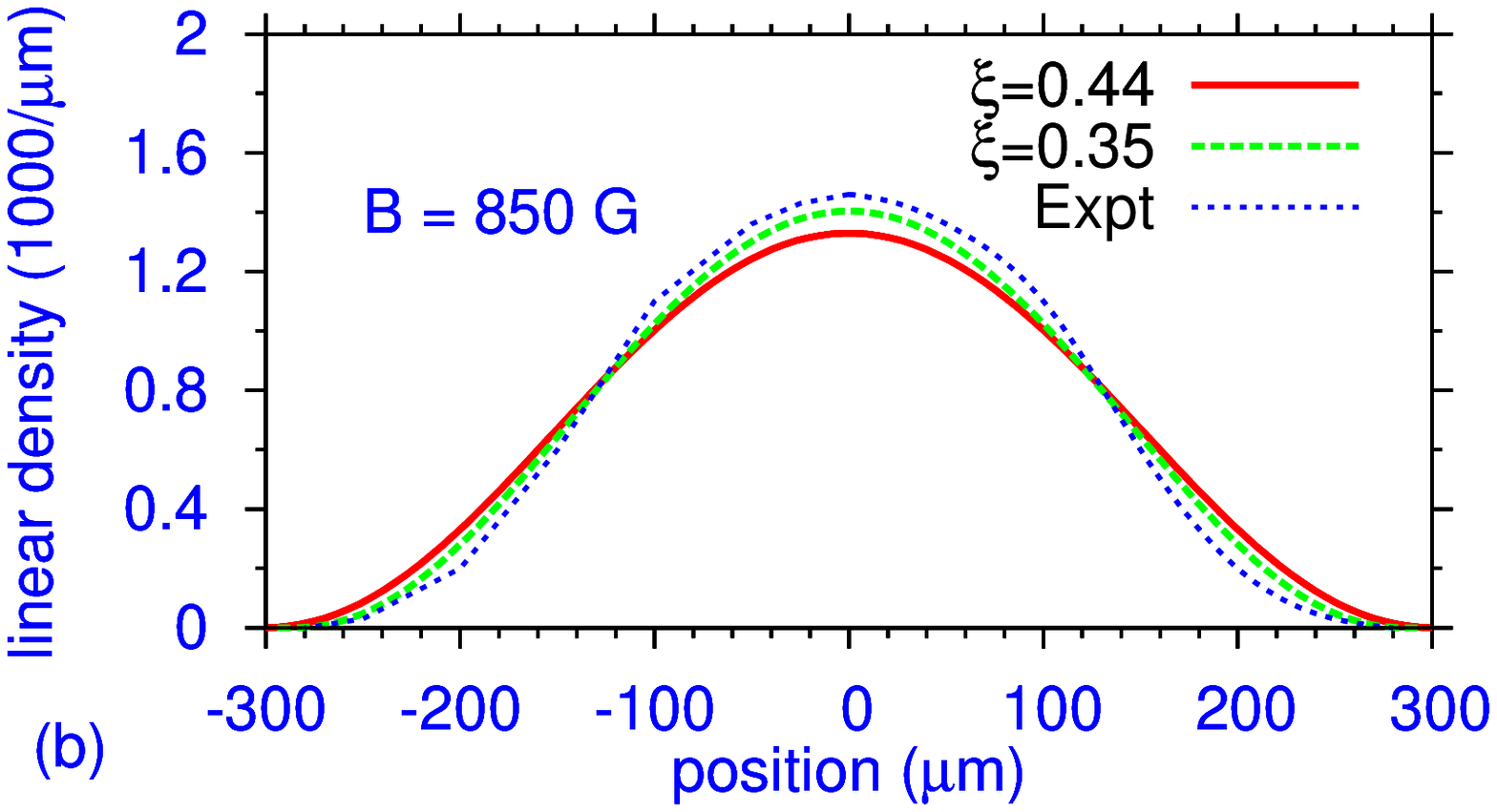}
\includegraphics[width=.49\linewidth]{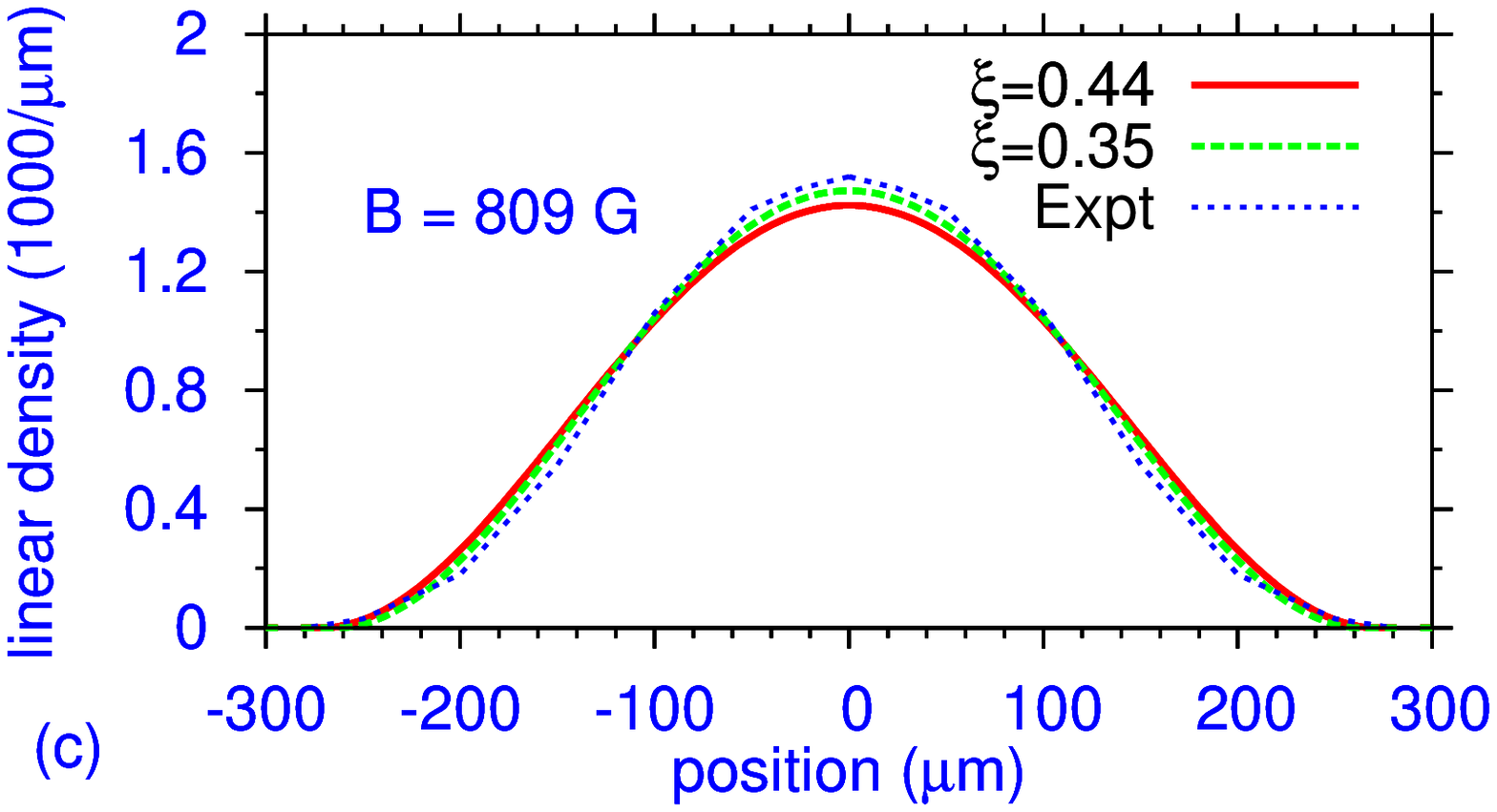}
\includegraphics[width=.49\linewidth]{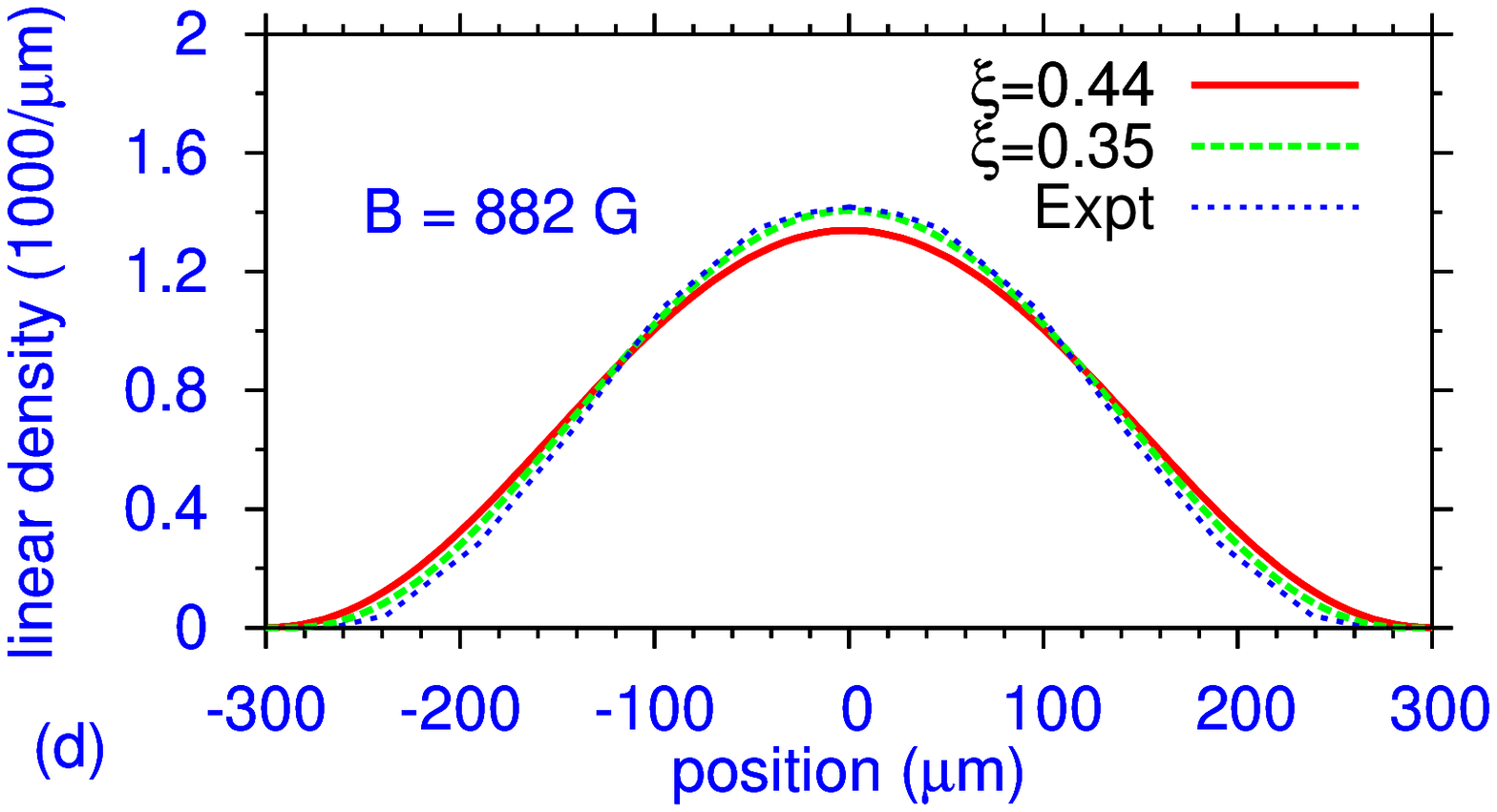}
\end{center}
\caption{(Color online) Linear density profile of the Fermi SF 
for magnetic field $B$  = (a) 764 G, (b) 850 G, (c) 809 G, and (d) 882 G, 
and for $N=4\times 10^5$ $^6$Li atoms, as 
calculated by using the DF equation for different $\xi$ 
and from experiment \cite{prl}.   
} \label{fig3}
\end{figure}

Next we present the results of linear density in Figs. \ref{fig3}
(a), (b), (c), and (d) for magnetic field $B=$  764 G 
(estimated $a=4.474\times 10^3a_0 $), 850 G 
($a=-2.817\times 10^4a_0$),
809 G (estimated $a=1.52 \times 10^4a_0$), 
and 882 G (estimated $a=-1.041\times 10^4a_0$), respectively, 
as calculated from the DF 
equation for different  $\xi$ and compare them with experiment.
(The present estimates for scattering length are slightly different from 
those presented in  \cite{perali}.)
From figure \ref{fig3} (a) we see that 
at $B=764$ G the linear density calculated with $\xi = 0.44$ and 0.35 are 
practically the same. This is because this value of $B$ is deep into the 
BEC region and away from unitarity and the result is not very sensitive 
to the value of $\xi$. The GP equation for dimers yields good result there \cite{prl}. 
In figure \ref{fig3} (b) we find that 
near  unitarity ($B=850$ G) the experimental result prefers a smaller value of 
$\xi$, e.g. $\xi=0.3$ over $\xi =0.44$ and 0.35 in agreement with 
\cite{rmp2,prl}.  
From Figs.  \ref{fig3} (c) and (d) we find that for $B=809$ G and 882 G, 
$\xi= 0.35$ should better explain the experimental 
trend. 
In view of the results reported in Figs. \ref{fig2} and \ref{fig3}, 
further experimental and theoretical studies 
are needed to see how universal the constant $\xi$ is and how sensitive 
it is to the applied trap, magnetic field and the number of atoms \cite{jltp}. 
The approximate solution of the Bogoliubov equation as reported in 
 \cite{perali}  leads to a slightly inferior agreement 
with these 
experimental data compared to the present model.

\begin{figure}
\begin{center}
\includegraphics[width=.67\linewidth]{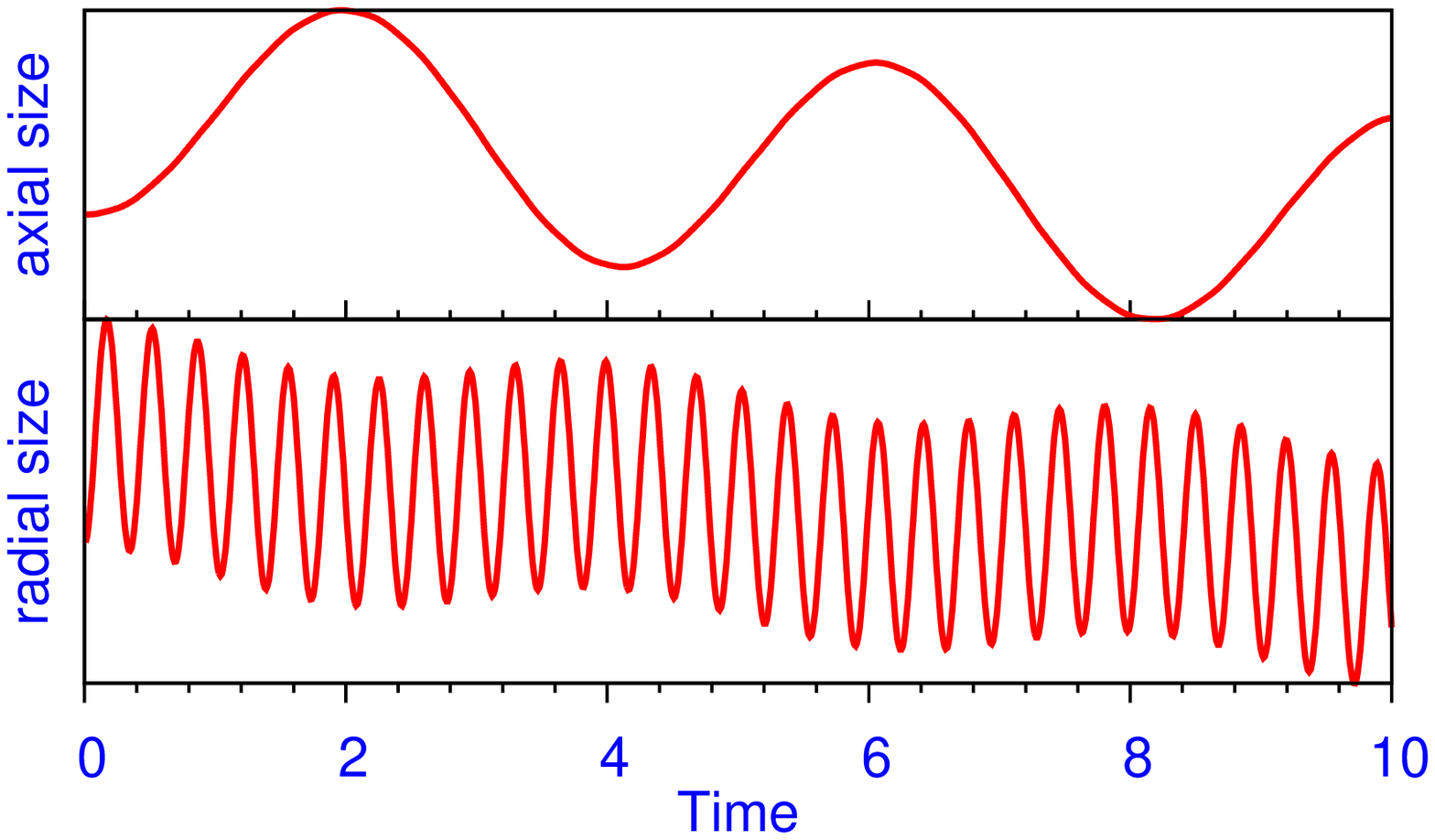}(a)
\includegraphics[width=.67\linewidth]{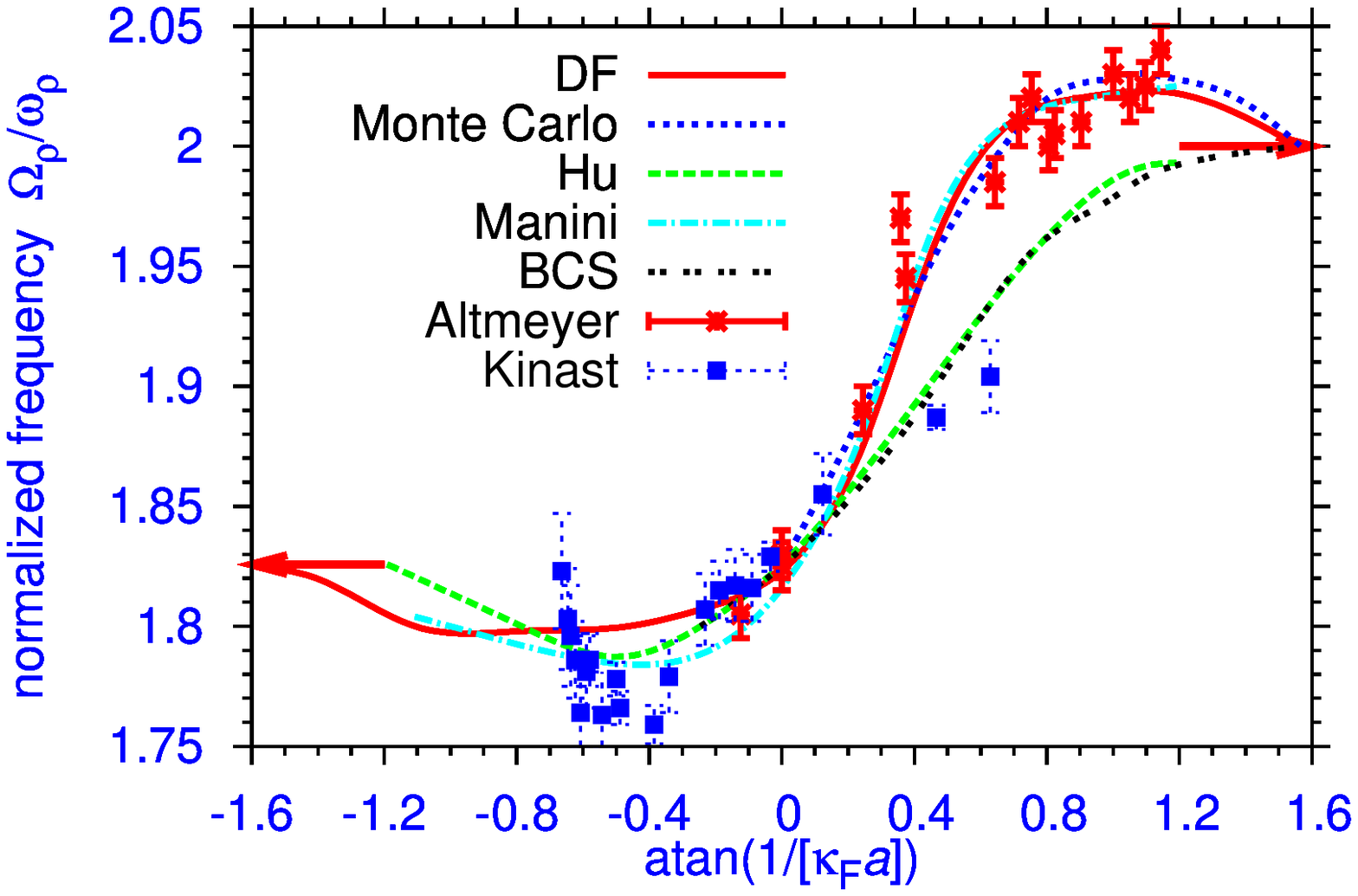}(b)
\includegraphics[width=.67\linewidth]{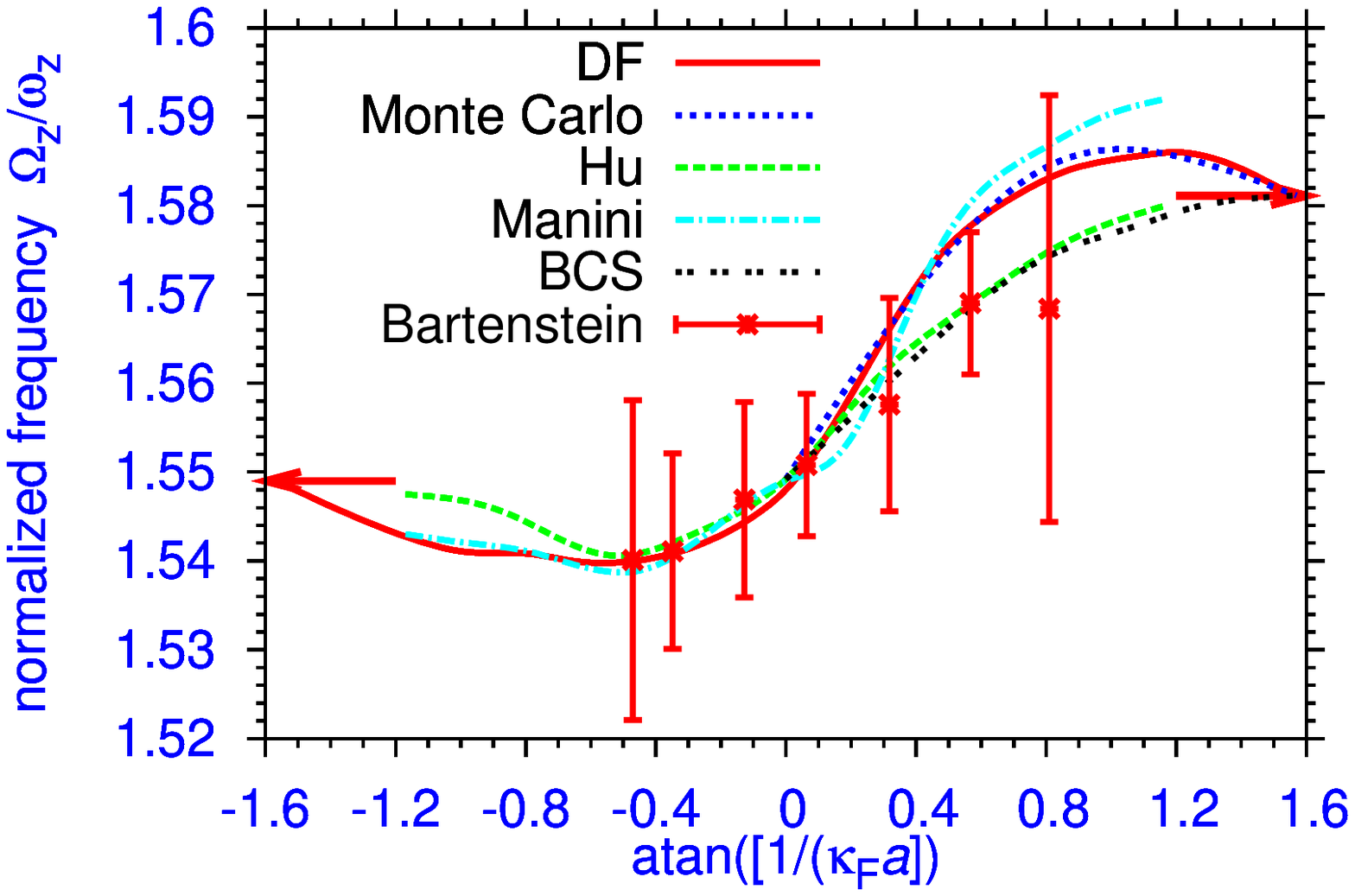}(c)
\end{center}

\caption{(Color online) (a) Typical small  
oscillation of axial and radial sizes 
of a cigar-shaped condensate after an initial  perturbation.
Normalized frequencies of (b) radial and (c) 
axial oscillation of a cigar-shaped Fermi SF vs. atan$([\kappa_Fa]^{-1})$ 
where $\kappa_F$ is the Fermi momentum in the axially symmetric trap. 
The arrows denote the analytic results of Cozzini and 
Stringari \cite{cozzini} in the BCS ($a\to -0$) and BEC ($a\to +0$) 
limits.  DF: present results with $\xi=0.44, \nu=1$;  
Monte Carlo: from  \cite{rmp2,altme,astra}; 
Hu: microscopic mean-field results \cite{hu}; 
Manini: numerical results from  the solution of a DF equation \cite{Sala};
BCS: from  \cite{rmp2,altme,astra}; 
Altmeyer: experiment \cite{altme};
Kinast: experiment \cite{kinast}; Bartenstein: experiment \cite{barten}.  
}
\label{fig4}
\end{figure}

\subsection{Oscillating Fermi SF}

\label{IIIb}

Now  we study the breathing oscillation of an elongated Fermi SF in the 
BCS-BEC crossover. The frequencies are obtained numerically from the 
oscillation \cite{exK2,altme,kinast,barten}
of a cigar-shaped 
Fermi SF. In our study we take the ratio 
of the trap frequencies $\omega_\rho/\omega_z=10$ and $N=40000$ atoms. 
Before we illustrate our results it is appropriate to present the 
analytical results for the problem suggested by Cozzini and Stringari 
\cite{cozzini} in 
the case of power-law dependence of the bulk chemical potential 
$\mu(n,a)$ as a function of $n: \mu(n,a)\sim n^\gamma$. The collective 
radial breathing mode frequency is given by $\Omega_\rho/\omega_\rho 
=\sqrt{2(1+\gamma)}$, and the collective axial breathing mode frequency 
is given by $\Omega_z/\omega_z
=\sqrt{(2+3\gamma)/(1+\gamma)}$. At unitarity and the BCS limit $\gamma =
2/3$, and 
consequently, $\Omega_\rho/\omega_\rho=\sqrt{10/3}\approx 1.82574 $ and 
$\Omega_z/\omega_z=\sqrt{12/5}\approx 1.54919$; at the BEC limit 
$\gamma=1$, and consequently, $\Omega_\rho/\omega_\rho=2$ and 
$\Omega_z/\omega_z=\sqrt{5/2}\approx  1.58114$.

The radial and axial frequencies of a cigar-shaped Fermi  SF  are calculated 
by studying the time evolution of a pre-formed condensate 
in a slightly altered trapping 
condition. The radial and axial sizes execute periodic 
oscillation of the type 
reported in figure  \ref{fig4} (a), from which the radial and axial frequencies 
are calculated reasonably accurately.
In the case of small breathing oscillations,
these frequencies are found to be 
independent of the perturbation that initiated the oscillation. 
The normalized frequencies $\Omega_\rho/\omega_\rho$ 
and $\Omega_z/\omega_z$  are also independent of $\omega_z$,  $\omega_\rho$
and $N$, provided that $N$ is large and the cigar-shaped condition 
is satisfied: $\omega_\rho>>\omega_z$.
The results for normalized radial and axial frequencies are plotted vs. 
$1/(\kappa_F a)$ in Figs. \ref{fig4} (b) and (c) for $\xi=0.44$, where 
$\kappa_F$ is the Fermi momentum,  and compared with 
different experiments and other calculations. (Unlike in 
Figs. \ref{fig2}  and \ref{fig3}, we could not find a noticable change in 
the frequecies as $\xi$ is changed from 0.44 to 0.35.)
In the axially symmetric 
trap the Fermi energy ${\cal E}_F$
is defined as ${\cal E}_F= (6N)^{1/3}\hbar 
(\omega_\rho^2\omega_z)^{1/3}$ and the Fermi 
momentum $\kappa_F$
is defined by $ {\cal E}_F=\hbar^2\kappa_F^2/m$. (Recall that 
$N$ is the number of molecules/pairs of mass $m$.)
The experimental radial frequencies of figure \ref{fig4}
(b) were taken from  \cite{altme,kinast} while the  experimental axial 
frequencies of figure \ref{fig4}
(c)
 were taken from  \cite{barten}.
In figure \ref{fig4} (b) we compare the radial frequencies with the Monte 
Carlo results of Astrakharchik {\it et al.} \cite{astra}
as quoted in  
\cite{rmp2,altme}, the
mean-field BCS Bogoliubov-de Gennes  
model calculation of Hu {\it et al.} \cite{hu}, and a numerical 
solution of a DF equation
 of Manini {\it et al.} \cite{Sala}. 
In figure \ref{fig4} (c) we compare the axial frequencies with the  Monte
Carlo results of Astrakharchik {\it et al.} \cite{astra}, the 
mean-field 
model calculation of Hu {\it et al.} \cite{hu}, and a 
numerical solution of a DF equation by Manini {\it et al.} \cite{Sala}. 
The numerical calculation of Manini {\it et al.} was performed with 
different cigar-shaped traps and different number of atoms from those 
used in the present study. Also, their gradient term in the DF equation 
was different from ours. The BCS and Monte Carlo calculations for the 
frequencies 
are performed \cite{astra}
by using the  BCS \cite{BCS} and Monte Carlo \cite{the1} equations of state
in a hydrodynamic approach.  The input for the calculation of frequencies 
  labeled Hu \cite{hu} and BCS \cite{astra} are both based on BCS 
mean-field approach and lead to very similar results. Qualitatively, all 
the theoretical results have the same trend. However, the curves labeled 
Hu and BCS do not have the over-shooting while approaching the BEC 
limit as in the experiment and the curves based on the DF approach and Monte 
Carlo equation of state. In the BEC-unitarity crossover the present frequencies 
are very close to the frequencies based on the Monte 
Carlo equation of state.

\section{Conclusion}

\label{IV}

We have suggested a generalized time-dependent Galilean-invariant
DF equation for a trapped Fermi SF 
valid in the full BCS-BEC crossover, which is equivalent to a quantum 
hydrodynamical equation. 
The equation of state has the correct 
model-independent 
limiting behaviors 
near unitarity and 
near the BCS \cite{lee}
and BEC \cite{huang}
limits as established by Lee, Yang, and Huang. 
While applied to the study of size and density profile of a 
cigar-shaped Fermi SF, the present formulation yields results 
in agreement with an experiment on $^6$Li atoms \cite{prl}. 
We also applied the present DF equation to the study of breathing 
frequencies of a
cigar-shaped trapped Fermi SF and the results are in good agreement
with experiment \cite{altme,kinast,barten}.

We also compare the present results with other theoretical findings 
\cite{BCS,Sala,perali,astra,hu}. The size and linear densities exhibited 
in Figs.  \ref{fig2} and \ref{fig3} are in reasonable agreement with 
those obtained from an approximate solution of the Bogoliubov-de Gennes 
equation \cite{perali}. The accurate solution of the Bogoliubov-de 
Gennes equation is numerically far more complicated than the solution of 
the the present DF equation and the present simulation yields slightly 
better agreement to the experimental data \cite{prl} 
than the approximate numerical 
solution of the Bogoliubov-de Gennes equation in  \cite{perali}. The 
present frequencies for axial and radial oscillations of a cigar-shaped 
Fermi SF are in excellent agreement with hydrodynamical results \cite{astra} 
based on 
a Monte Carlo equation of state \cite{the1} and the precise experiment of Altmeyer 
{\it et al.} \cite{altme}
but are in disagreement with a 
mean-field BCS equation of state \cite{BCS} and the experiment of Kinast 
{\it et al.} \cite{kinast} except near unitarity. The data of  Kinast
{\it et al.} \cite{kinast} away from unitarity seems to be inaccurate and 
the overshooting of the frequencies in Figs. \ref{fig4} 
above the Cozzini-Stringari limit \cite{cozzini}
in the BEC side of crossover is appropriate.
More precise experiments can determine 
which of the theories are more realistic. The experiments are easier in 
the BEC-unitarity region and in this region there is some disagreement 
between two experiments \cite{altme,kinast} for the radial frequencies 
reported in figure \ref{fig4} (b), and the large experimental error in the 
axial frequencies \cite{barten} does not allow us to choose  between the 
mean-field calculations \cite{BCS,hu} on the one hand and the present 
and Monte Carlo calculations \cite{astra} on the other hand.  
More 
accurate experiments are needed to resolve the discrepancy.

Although the present approach yields good results for many experimental 
observations, it has its limitation too. For many dynamical problems, 
such as in rotation and vortex creation, breaking of the weakly bound 
pairs in the BCS-unitarity crossover may play a vital role in the 
phenomenology of the Fermi super-fluid. 
{\color{Blue}At sufficiently small negative $\kappa_Fa$, 
both the Innsbruck and Duke groups 
observed a breakdown of hydrodynamics in the radial breathing mode, 
presumably because the excitation frequency of the collective mode 
exceeded the pairing energy in the weakly interacting regime \cite{kinast}.}
Pair breaking is not permitted in 
the present DF formulation, which may lead to inappropriate result in these 
problems. No such problem should appear in the application of the present 
formulation to a description of density profile, sizes, and frequencies of 
small oscillation as in this paper.

\ack

FAPESP and CNPq (Brazil) provided partial support.

\end{document}